\documentclass[conference,10pt]{IEEEtran}
\usepackage{epsfig,rotating,setspace,latexsym,amsmath,epsf,amssymb,amsfonts,bm,theorem,subfigure,epstopdf}
\usepackage{cite,authblk}
\usepackage{bbm}
\usepackage{algorithm}
\usepackage[noend]{algpseudocode}

\algrenewcommand\algorithmicforall{\textbf{foreach}}
\algrenewcommand\algorithmicindent{.8em}
\usepackage{color}
\usepackage{mathtools}

\newtheorem{theorem}{Theorem}

\newenvironment{Proof}[1]{\medskip\par\noindent{\bf Proof:\,}\,#1}{{\mbox{\,$\blacksquare$}\par}}
\IEEEoverridecommandlockouts
\allowdisplaybreaks

\begin{document}

\title{Maximizing Information Freshness in Caching Systems with Limited Cache Storage Capacity \thanks{This work was supported by NSF Grants CCF 17-13977 and ECCS 18-07348.}}

\author{Melih Bastopcu \qquad Sennur Ulukus\\
	\normalsize Department of Electrical and Computer Engineering\\
	\normalsize University of Maryland, College Park, MD 20742\\
	\normalsize \emph{bastopcu@umd.edu} \qquad \emph{ulukus@umd.edu}}

\maketitle

\begin{abstract}	
We consider a cache updating system with a source, a cache with limited storage capacity and a user. There are $n$ files. The source keeps the freshest versions of the files which are updated with known rates. The cache gets fresh files from the source, but it can only store the latest downloaded versions of $K$ files where $K\leq n$. The user gets the files either from the cache or from the source. If the user gets the files from the cache, the received files might be outdated depending on the file status at the source. If the user gets the files directly from the source, then the received files are always fresh, but the extra transmission times between the source and the user decreases the freshness at the user. Thus, we study the trade-off between storing the files at the cache and directly obtaining the files from the source at the expense of additional transmission times. We find analytical expressions for the average freshness of the files at the user for both of these scenarios. Then, we find the optimal caching status for each file (i.e., whether to store the file at the cache or not) and the corresponding file update rates at the cache to maximize the overall freshness at the user. We observe that when the total update rate of the cache is high, caching files improves the freshness at the user. However, when the total update rate of the cache is low, the optimal policy for the user is to obtain the frequently changing files and the files that have relatively small transmission times directly from the source.
\end{abstract}

\section{Introduction}

Time sensitive information has become ever more important especially with emerging technologies such as autonomous driving, augmented reality, social networking, high-frequency automated trading, online gaming, and so on. Age of information has been introduced to measure the timeliness of information in communication networks. Age of information has been widely studied in the context of web crawling, queueing networks, caching systems, remote estimation, energy harvesting systems, scheduling in networks, and so on \cite{Cho03, Kolobov19a, Kaul12a, Costa14, Soysal19, Gao12, Yates17b, Kam17b, Zhong18c, Zhang18, Tang19, Yang19a, Sun17b, Arafa20a, Bastopcu18, Buyukates18c, Bastopcu20a, Bastopcu20d}.
 
\begin{figure}[t]
	\centering  \includegraphics[width=0.83\columnwidth]{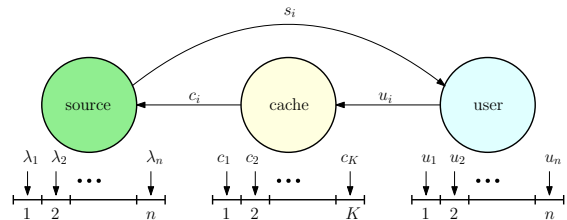}
	\caption{A cache updating system with a source, a cache and a user.}
	\label{fig:model}
	\vspace{-5mm}
\end{figure}

In this work, we consider a cache updating system that consists of a source, a cache with limited cache (i.e., storage) capacity and a user as shown in Fig.~\ref{fig:model}. In this system, the source keeps the freshest versions of all the files that are refreshed with known rates $\lambda_i$. The cache gets the freshest versions of the files from the source, but its cache capacity is limited, i.e., it can only store the freshest versions of $K$ files where $K\leq n$. The user gets files either from the cache or from the source. If the user gets a file from the cache, the updated file at the user might still be outdated depending on the file status at the source. If the user gets a file directly from the source, the received file is always fresh. However, as the channel between the user and the source is not perfect, there is a file transmission time which decreases the freshness at the user. Thus, in this paper, we study the trade-off between storing the files at the cache to decrease the file transmission times versus directly obtaining the fresh files from the source at the expense of higher transmission times. Our aim is to find the optimal caching status for each file (i.e., whether to store the file at the cache or not) and the corresponding optimal file update rates at the cache.                   

References that are most closely related to our work are \cite{Yates17b} and \cite{Bastopcu20d}. Reference \cite{Yates17b} considers a model where a resource constrained remote server wants to keep the items at a local cache as fresh as possible. Reference \cite{Yates17b} shows that the update rates of the files should be chosen proportional to the square roots of their popularity indices. Different from \cite{Yates17b} where the freshness of the local cache is considered, we consider the freshness at the end-user. Furthermore, the freshness metric that we use in this paper is different than the traditional age metric used in \cite{Yates17b}, and hence our overall work is distinct compared to \cite{Yates17b}. In comparison to our earlier work in \cite{Bastopcu20d}, here, we consider a cache with limited caching capacity, and we study the trade-off between storing the files at the cache and obtaining the files directly from the source.   

In this paper, we find an analytical expression for the average freshness of the files at the user when the files are cached and not cached. We impose a total update rate constraint for the cache due to limited nature of resources. We find the optimal caching status for each file and the corresponding optimal file update rates at the cache. We observe that due to binary nature of file caching status, the optimization problem is NP-hard. However, for a given set of caching status of the files, the problem becomes a convex optimization problem in terms of the file update rates at the cache. For a given set of caching status of the files, the optimal rate allocation policy at the cache is a \emph{threshold policy} where the rapidly changing files at the source may not be updated. We observe that when the total update rate of the cache is high, storing files at the cache improves the freshness of the user. However, when the total update rate of the cache is low, it is optimal for the user to obtain the rapidly changing files and the files that have relatively small transmission times directly from the source. 

\section{System Model} \label{sect:system_model}

We consider an information updating system where there is a source, a cache and a user as shown in Fig.~\ref{fig:model}. The source keeps the freshest version of $n$ files which are updated with exponential inter-arrival times with rate $\lambda_i$. The file updates at the source are independent of each other. The cache gets fresh files from the source, but it may store only $K$ files where $K=1,\dots,n$. We assume that the channel between the source and the cache is perfect and the transmission times are negligible. Thus, if the cache requests an update for a stored file, it receives the file from the source right away. We model the inter-update request times for the $i$th file at the cache as exponential with rate $c_i$. The cache is subject to a total update rate constraint, i.e., $\sum_{i=1}^{n}c_i \leq C$ as in \cite{Yates17b, Bastopcu20d}. 

The inter-update request times of the user for the $i$th file are exponential with rate $u_i$. The channel between the user and the cache is also assumed to be perfect and the transmission times are negligible. Thus, if the requested file is stored at the cache, the user gets the stored file at the cache right away. If the user requests a file which is not cached, the cache forwards the file update request from the user to the source. Since the cache only forwards the user requests for uncached files (i.e., without creating requests of its own), $c_i>u_i$ is not possible, and we have $c_i\leq u_i$. For uncached files, the file update requests at the cache are fully synchronized with the user's requests which means that when the user requests an update for an uncached file, this request reaches the source immediately if the cache forwards it. Thus, for each file update request of the user for the uncached file $i$, the cache forwards the request to the source with probability $p_i = \frac{c_i}{u_i}$. From \cite[Thm.~13.6]{Yates14}, the effective inter-update request times of the user for an uncached file are exponential with rates $c_i$. We assume that the channel between the source and the user is  imperfect and the transmission time for the $i$th file is exponential with rate $s_i$.      

We note that each file at the source is always \textit{fresh}. However, when a file is updated at the source, the stored versions of the same file at the cache and at the user become \textit{outdated}. When the cache gets an update for an outdated file, the updated file in the cache becomes \textit{fresh} again until the next update arrival at the source. The user gets files either from the cache or from the source. If the user gets a file from the cache, it will receive the file update immediately, but the received file can be outdated if the file at the cache is not fresh. If the user gets a file directly from the source, the received file is always fresh, but the transmission takes time. We note that since the cache and the user are unaware of the file updates at the source, they do not know whether they have the freshest versions of the files or not. Thus, they may still unknowingly request an update even though they have the freshest version of a file. 

\begin{figure}[t]
	\begin{center}
		\subfigure[]{%
			\includegraphics[width=0.49\linewidth]{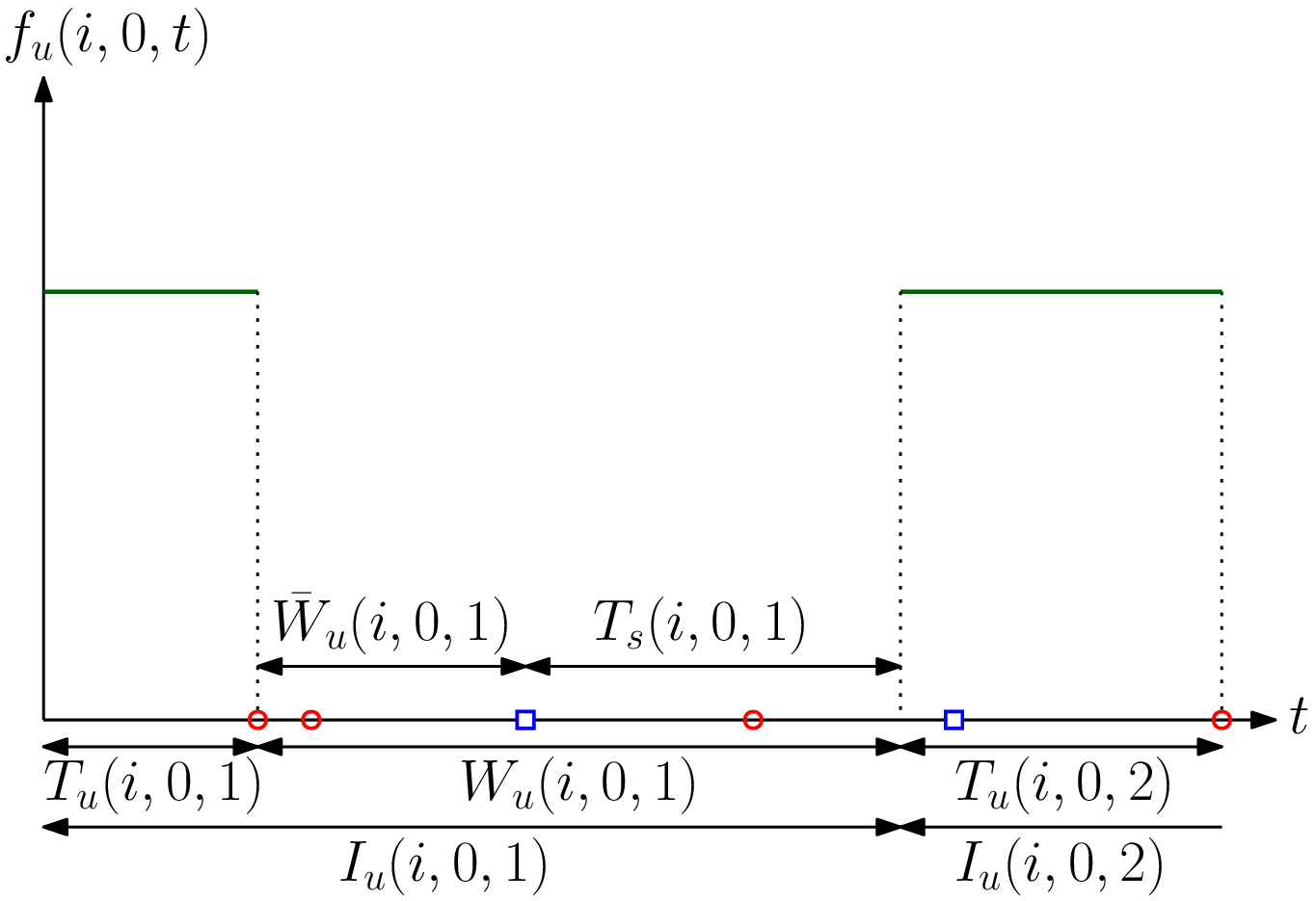}}
		\subfigure[]{%
			\includegraphics[width=0.49\linewidth]{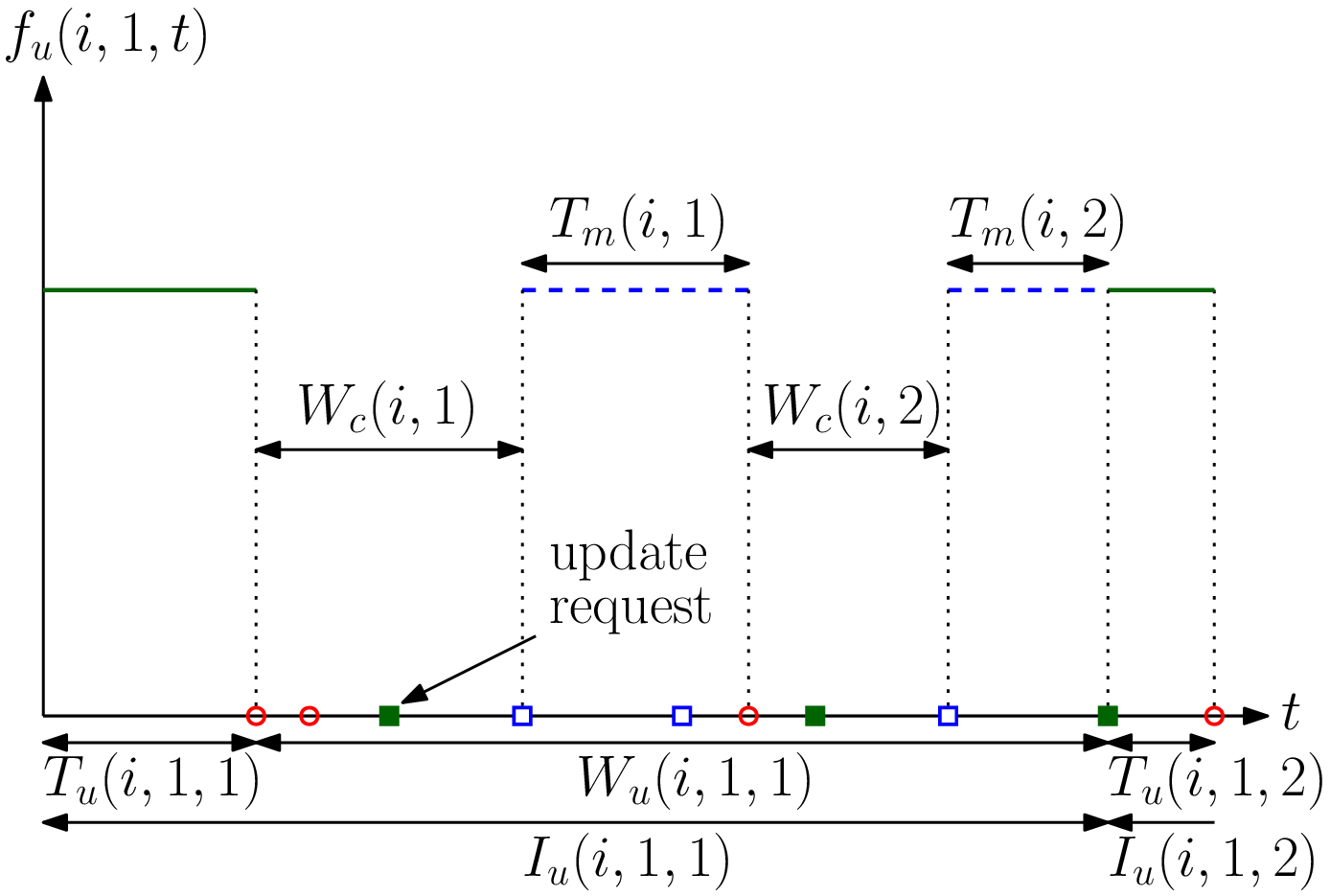}}
	\end{center}
	\caption{Sample evolution of the freshness of the $i$th file at the user when the $i$th file is (a) not cached and (b) cached. Red circles represent the update arrivals at the source, blue squares represent the update requests from the cache, and green filled squares represent the update requests from the user.}
	\label{Fig:age_evol}
	\vspace{-5mm}
\end{figure} 

We use $k_i$ which is a binary variable to indicate the caching status of the $i$th file, i.e., $k_i  = 1$ when the $i$th file is cached and $k_i = 0$ when it is not cached. We define $f_{u}(i,k_i,t)$ as the freshness function of the $i$th file at the user as,
\begin{align}
f_{u}(i,k_i,t) = \begin{cases} 
1, & \text{if the $i$th file is fresh at time $t$}, \\
0, & \text{otherwise,}
\end{cases}
\end{align}
where the instantaneous freshness function is a binary function taking values of fresh, ``$1$'', or not fresh, ``$0$'', at any time $t$. A sample $f_{u}(i,k_i,t)$ is shown in Fig.~\ref{Fig:age_evol}(a) when $k_i=0$ and in Fig.~\ref{Fig:age_evol}(b) when $k_i = 1$. 

File updates that replace an outdated version of the file with the freshest one are denoted as \textit{successful} updates. We define the time interval between the $j$th and the $(j+1)$th successful updates for the $i$th file at the user as the $j$th update cycle and denote it by $I_{u}(i,k_i,j)$. We denote the time duration when the $i$th file at the user is fresh during the $j$th update cycle as $T_{u}(i,k_i,j)$. We denote $f_c(i,t)$ as the freshness function of the $i$th file at the cache. Similarly, update cycles and duration of freshness at the cache are denoted by $I_{c}(i,j)$ and $T_c(i,j)$. Then, we denote $F_{u}(i, 1)$ (resp. $F_{u}(i, 0)$) as the long term average freshness of the $i$th file at the user when the file is cached (resp. when the file is not cached), i.e., $k_i = 1$ (resp. $k_i = 0$). $F_{u}(i, k_i)$ is equal to
\begin{align}
F_{u}(i,k_i) = \lim_{T\rightarrow\infty} \frac{1}{T} \int_{0}^{T} f_{u}(i,k_i,t)dt.    
\end{align}
Similar to \cite{Cho03}, we have
\begin{align*}
F_{u}(i,k_i) = \lim\limits_{T\rightarrow \infty}\frac{N}{T}\left(\frac{1}{N}\sum_{j=1}^{N}T_{u}(i,k_i,j)\right) = \frac{\mathbb{E}[T_{u}(i,k_i)]}{\mathbb{E}[I_{u}(i,k_i)]},
\end{align*}
where $N$ is the number of update cycles in time duration $T$. We define the total freshness over all files at the user $F_u$ as
\begin{align}\label{total_freshness_user_new}
F_u = \sum_{i=1}^{n}k_i F_{u}(i,1)+(1-k_i) F_{u}(i,0).    
\end{align}

Our aim is to find the optimal file caching status $k_i$, and the corresponding file update rates at the cache $c_i$ for $i=1,\dots,n$, such that the total average freshness of the user $F_u$ is maximized while satisfying the constraints on the cache capacity, i.e., $\sum_{i=1}^{n}k_i\leq K$, the total update rate of the cache, $\sum_{i=1}^{n}c_i\leq C$, and the feasibility constraints i.e., $c_i\leq u_i$ for uncached files (for files with $k_i =0$). Thus, our problem is,
\begin{align}
\label{problem1}
\max_{\{k_i, c_i \}}  \quad &  F_u\nonumber \\
\mbox{s.t.} \quad & \sum_{i=1}^{n}k_i\leq K \nonumber \\
\quad & \sum_{i=1}^{n}c_i\leq C \nonumber \\
\quad &(1-k_i)c_i\leq u_i,\quad \qquad i=1,\dots,n \nonumber \\
\quad & c_i\geq 0, \quad k_i\in\{0,1\},\quad i=1,\dots,n.
\end{align} 

In the following section, we find the long term average freshness of the $i$th file at the user $F_{u}(i,k_i)$ when the $i$th file is cached and when it is not cached. Once we find $F_u(i, k_i)$, this will determine the objective function of (\ref{problem1}) via (\ref{total_freshness_user_new}).

\section{Average Freshness Analysis} \label{Sec:Average_freshness} 

In this section, we find the long term average freshness for the $i$th file at the user $F_u(i, k_i)$ for $k_i\in\{0,1\}$. In the following theorem, we first find the long term average freshness of the $i$th file at the user when the $i$th file is cached.

\begin{theorem}\label{thm1}
	If the $i$th file is cached, the long term average freshness of the $i$th file at the user $F_u(i,1)$ is equal to 
	\begin{align}
	F_{u}(i,1) = \frac{\mathbb{E}[T_{u}(i,1)]}{\mathbb{E}[I_{u}(i,1)]} = \frac{u_i}{u_i+\lambda_i}\frac{c_i}{c_i+\lambda_i}.\label{freshness_user_cached}
	\end{align} 
\end{theorem}

The proof of the Theorem~\ref{thm1} follows from \cite[Section~III]{Bastopcu20d}. Since the user gets fresh files more frequently from the cache for higher values of $c_i$, the freshness of the $i$th file at the user $F_u(i,1)$ in (\ref{freshness_user_cached}) increases with $c_i$. In addition, $F_u(i,1)$ in (\ref{freshness_user_cached}) is a concave function of $c_i$. If the user was directly connected to the source, freshness of the $i$th file at the user would be equal to $ \frac{u_i}{u_i+\lambda_i}$ as in \cite{Bastopcu20d}. However, as the user is connected to the source via the cache, the freshness experienced by the user proportionally decreases with the freshness of the cache which is $\frac{c_i}{c_i+\lambda_i}$. Note that $\frac{c_i}{c_i+\lambda_i}<1$ for all $c_i$. 

Next, we find the long term average freshness of the $i$th file at the user $F_u(i,0)$ when the $i$th file is not cached.  

\begin{theorem}\label{thm2}
	If the $i$th file is not cached, the long term average freshness of the $i$th file at the user $F_u(i,0)$ is equal to 
	\begin{align}
	F_{u}(i,0) = \frac{\mathbb{E}[T_{u}(i,0)]}{\mathbb{E}[I_{u}(i,0)]} = \frac{c_i}{c_i+\lambda_i+\frac{ c_i \lambda_i }{s_i}}.\label{freshness_user_uncached}
	\end{align} 
\end{theorem}
\begin{Proof}
When the $i$th file at the user becomes fresh, the time until the next file update arrival at the source is still exponentially distributed with rate $\lambda_i$ due to the memoryless property of the exponential distribution. Thus, $\mathbb{E}[T_u(i,0)] = \frac{1}{\lambda_i}$.

After the $i$th file is updated at the source, the stored version of the $i$th file at the user becomes outdated, i.e., the instantaneous freshness function $f_u(i,0,t)$ becomes $0$ again. We denote the time interval until the source gets a file update request for the $i$th file after the file at the user becomes outdated as $\bar{W}_u(i,0)$ which is exponentially distributed with rate $c_i$ as discussed in Section~\ref{sect:system_model}. After receiving the file update request from the user, the source sends the $i$th file directly to the user. If the $i$th file at the source is updated during a file transfer, then the file transfer is interrupted and the fresh file is sent until the freshest version of the $i$th file is successfully transmitted to the user. We denote the total transmission time for the $i$th file as $T_s(i,0)$. Due to \cite[Prob. 9.4.1]{Yates14}, $T_s(i,0)$ is also exponentially distributed with rate $s_i$. Thus, we have $\mathbb{E}[T_s(i,0)] = \frac{1}{s_i}$. We denote the time interval when the $i$th file at the user is outdated during the $j$th update cycle as $W_u(i,0,j)$, i.e., $W_u(i,0,j) = I_u(i,0,j)- T_u(i,0,j)$, which is also equal to $ W_u(i,0,j) = \bar{W}_u(i,0,j)+T_s(i,0,j)$. We denote the typical random variables for $W_u(i,0,j)$ and $I_u(i,0,j)$ as $W_u(i,0)$ and $I_u(i,0)$, respectively. Then, we have $ \mathbb{E}[W_u(i,0)] = \mathbb{E}[\bar{W}_u(i,0)]+\mathbb{E}[T_s(i,0)] = \frac{1}{c_i}+\frac{1}{s_i}$ and
\begin{align*}
\mathbb{E}[I_u(i,0)] = \mathbb{E}[T_u(i,0)] +\mathbb{E}[W_u(i,0)] = \frac{1}{\lambda_i}+\frac{1}{c_i}+\frac{1}{s_i}.
\end{align*}
Thus, we get $F_{u}(i,0)$ in (\ref{freshness_user_uncached}) by using $F_{u}(i,0) = \frac{\mathbb{E}[T_{u}(i,0)]}{\mathbb{E}[I_{u}(i,0)]}$.
\end{Proof}

We note that $F_u(i,0)$ in (\ref{freshness_user_uncached}) is an increasing function of $c_i$ and also is concave in $c_i$. When the user gets a file from the source directly, the received file is always fresh, but due to the transmission time between the source and the user, the average time that the $i$th file is outdated at the user increases. Thus, the freshness of the $i$th file at the user $F_u(i,0)$ in (\ref{freshness_user_uncached}) increases with $s_i$. 
Further, $F_u(i,1) > F_u(i,0)$ implies that $\frac{c_i}{c_i+\lambda_i}>\frac{s_i}{u_i}$. In other words, if the file update rate of the $i$th file at the cache $c_i$ is high enough, it is better to cache file $i$. However, if file $i$ is updated too frequently at the source, i.e., $\lambda_i$ is too large, or file $i$ has small transmission times, i.e., $s_i$ is too high, then it is better to get the file from the source. Thus, there is a trade-off: If a file is stored at the cache, this enables the user to obtain the file more quickly, but the received file might be outdated. On the other hand, if the user gets the file directly from the source, the file will always be fresh, but the file transmission time decreases the freshness at the user.     

\section{Freshness Maximization}\label{sect:gen_soln}

In this section, we solve the optimization problem in (\ref{problem1}). Using $F_u(i,k_i)$ in (\ref{freshness_user_cached}) and (\ref{freshness_user_uncached}) and $F_u$ in (\ref{total_freshness_user_new}), we rewrite the freshness maximization problem in (\ref{problem1}) as 
\begin{align}
\label{problem1_mod}
\max_{\{k_i, c_i \}}  \quad &  \sum_{i=1}^{n}k_i \frac{u_i}{u_i+\lambda_i}\frac{c_i}{c_i+\lambda_i}+(1-k_i) \frac{c_i}{c_i+\lambda_i+\frac{ c_i \lambda_i }{s_i}}\nonumber \\
\mbox{s.t.} \quad & \sum_{i=1}^{n}k_i\leq K \nonumber \\
\quad & \sum_{i=1}^{n}c_i\leq C \nonumber \\
\quad & (1-k_i)c_i\leq u_i,\quad \qquad i=1,\dots,n \nonumber \\
\quad & c_i\geq 0, \quad k_i\in\{0,1\},\quad i=1,\dots,n.
\end{align}

In order to solve the optimization problem in (\ref{problem1_mod}), we need to determine the optimal caching status for each file $k_i$ and find the optimal file update rates at the cache $c_i$. We note that the optimization problem in (\ref{problem1_mod}) is NP-hard due to the presence of binary variables $k_i$. However, for a given $(k_1,k_2,\dots,k_n)$ tuple, (\ref{problem1_mod}) becomes a convex optimization problem in $c_i$. Thus, the optimal solution can be found by searching over all possible $(k_1,k_2,\dots,k_n)$ tuples and finding the corresponding optimal $c_i$ values for each $(k_1,k_2,\dots,k_n)$ tuple.      

Next, for a given set of $(k_1,k_2,\dots,k_n)$ values, we find the corresponding optimal $c_i$ values. For that, we introduce the Lagrangian function \cite{Boyd04} for (\ref{problem1_mod}) as 
\begin{align*}
\mathcal{L} =&-\sum_{i=1}^{n}k_i \frac{u_i}{u_i+\lambda_i}\frac{c_i}{c_i+\lambda_i}+(1-k_i) \frac{c_i}{c_i+\lambda_i+\frac{ c_i \lambda_i }{s_i}}\nonumber \\
& +\beta\left(  \sum_{i=1}^{n}c_i- C\right)+\sum_{i=1}^{n} \eta_i((1-k_i)c_i-u_i)- \sum_{i=1}^{n}\nu_i c_i,
\end{align*}
where $\beta\geq 0$, $ \eta_i\geq 0$ and $\nu_i\geq0$. The KKT conditions are 
\begin{align}
\frac{\partial \mathcal{L}}{\partial c_i} &= -\frac{u_i}{u_i+\lambda_i}\frac{\lambda_i}{\left(c_i+\lambda_i\right)^2} +\beta-\nu_i = 0, \label{KKT1}
\end{align}
for all $i$ with $k_i = 1$, and 
\begin{align}
\frac{\partial \mathcal{L}}{\partial c_i} &= -\frac{\lambda_i}{\left(c_i+\lambda_i+\frac{c_i \lambda_i}{s_i}\right)^2} +\beta+\eta_i-\nu_i = 0,\label{KKT2}
\end{align}
for all $i$ with $k_i = 0$. Complementary slackness conditions are
\begin{align}
\beta\left(  \sum_{i=1}^{n}c_i- C\right) &= 0, \label{CS1}\\
\eta_i((1-k_i)c_i-u_i) &= 0, \label{CS3}\\
\nu_i c_i &= 0.\label{CS2}
\end{align}

For given $k_i$s with $k_i=1$, we rewrite (\ref{KKT1}) as 
\begin{align}
(c_i+\lambda_i)^2 = \frac{1}{\beta-\nu_i}\frac{u_i\lambda_i}{u_i+\lambda_i}.
\end{align}
If $c_i>0$, we have $ \nu_i = 0$ from (\ref{CS2}). Thus, we have
\begin{align}\label{soln_c_i_1}
c_i = \left(\frac{1}{\sqrt{\beta}} \sqrt{\frac{u_i \lambda_i}{u_i+\lambda_i}}-\lambda_i\right)^+,
\end{align}
for all $i$ with $k_i=1$, where $(x)^+ = \max(x,0)$. Similarly, for given $k_i$s with $k_i = 0 $, we rewrite (\ref{KKT2}) as 
\begin{align}
\left(c_i+\lambda_i+\frac{c_i \lambda_i}{s_i}\right)^2 = \frac{\lambda_i}{\beta+\eta_i-\nu_i}.
\end{align}
If $c_i>0$, we have $ \nu_i = 0$ from (\ref{CS2}). Furthermore, if $ c_i<u_i$, then we have $ \eta_i = 0$ from (\ref{CS3}). Otherwise, we have $ c_i = u_i$ and $ \eta_i \geq 0$ from (\ref{CS3}). Thus, we have
\begin{align}\label{soln_c_i_0}
c_i =\min \left(\frac{s_i}{s_i+\lambda_i}\left(\sqrt{\frac{\lambda_i}{\beta}}-\lambda_i \right)^+, u_i\right), 
\end{align}
for all $i$ with $k_i=0$.

Note that $c_i>0$ in (\ref{soln_c_i_1}) requires $ \frac{1}{\lambda_i}\frac{u_i}{u_i+\lambda_i}>\beta$ which also implies that if $ \frac{1}{\lambda_i}\frac{u_i}{u_i+\lambda_i}\leq \beta$, then we must have $c_i=0$. Similarly, $c_i>0$ in (\ref{soln_c_i_0}) requires $ \frac{1}{\lambda_i}>\beta$ which also implies that if $ \frac{1}{\lambda_i}\leq \beta$, then we must have $c_i=0$. Thus, for given $k_i$s, we observe that the optimal rate allocation policy for the cache is a \emph{threshold policy} in which the optimal update rates are equal to zero when the file update rates $\lambda_i$s are too large, i.e., when the files are updated too frequently at the source. In the optimal policy, the total update rate constraint for the cache, i.e., $ \sum_{i=1}^{n}c_{i}\leq C$, should be satisfied with equality as the objective function in (\ref{problem1_mod}) is an increasing function of $c_{i}$.

For given $k_i$s and $u_i$s, we define $\phi_i$ as
\begin{align} \label{phi-i-defn}
\phi_i =  \begin{cases} 
\frac{1}{\lambda_i}\frac{u_i}{u_i+\lambda_i}, & \text{if $k_i=1$}, \\
\frac{1}{\lambda_i}, & \text{if $k_i=0$}.
\end{cases}
\end{align}
Similar to \cite[Lemma~3]{Bastopcu20d}, for given $k_i$s and $u_i$s, if $c_i>0$ for some $i$, then we have $c_j> 0$ for all $j$ with $\phi_j\geq \phi_i$.

Next, for a given set of $k_i$s and $u_i$s, we find the optimal $c_i$s. First, we obtain $\phi_i$ from (\ref{phi-i-defn}). We initially assume that $c_i< u_i$ for all $i$ with $k_i =0$, i.e., $c_i$ in (\ref{soln_c_i_0}) is equal to $\frac{s_i}{s_i+\lambda_i}\left(\sqrt{\frac{\lambda_i}{\beta}}-\lambda_i \right)^+$. Then, we rewrite (\ref{soln_c_i_1}) and (\ref{soln_c_i_0}) as
\begin{align} \label{policy_finder}
c_i =  \begin{cases} 
\frac{\lambda_i}{\sqrt{\beta}}\left(\sqrt{\phi_i} -\sqrt{\beta}\right)^+, &\text{if $k_i=1$}, \\
\frac{s_i}{s_i+\lambda_i}\frac{\lambda_i}{\sqrt{\beta}}\left(\sqrt{\phi_i} -\sqrt{\beta}\right)^+, & \text{if $k_i=0$}.
\end{cases}
\end{align}

As we discussed earlier, in the optimal policy, we must have $\sum_{i=1}^{n}c_i = C$. Similar to the solution method in \cite{Bastopcu20d}, we solve $\sum_{i=1}^{n}c_i = C$ for $\beta$ by assuming that $\phi_i\geq\beta$ for all $i$, i.e., by ignoring $(\cdot)^+$ in (\ref{policy_finder}). Then, we compare the smallest $\phi_i$ with $\beta$. If the smallest $\phi_i$ is larger than or equal to $\beta$, it implies that $c_i>0$ for all $i$ as we assumed before, and we have obtained $c_i$ values for given $k_i$s. If the smallest $\phi_i$ is smaller than $\beta$, it implies that the corresponding $c_i$ was negative and it must be chosen as zero. In this case, we choose $c_i=0$ for the smallest $\phi_i$. Then, we repeat this process again until the smallest $\phi_i$ among the remaining $c_i$s satisfies $\phi_i\geq \beta$. 

Finally, when we find all $c_i$ values, we go back to our initial assumption which is $c_i< u_i$ for all $i$ with $k_i =0$ and check whether it holds or not. We define the set $S = \{i|c_i>u_i, k_i = 0\}$. If we have $c_i< u_i$ for all $i$ with $k_i =0$, i.e., when $S$ is empty, then we obtain the optimal $c_i$ values. If we have $c_i> u_i$ for some $i$ with $k_i =0$, then in the optimal policy, we have $c_i = u_i$ for all $i\in S$. Then, for remaining $c_i$s with $i\not\in S$, we repeat this process again with the remaining total update rate, i.e., $C-\sum_{i\in S}u_i$, until we have $c_i\leq u_i$ for all $i$ with $k_i = 0$.  

\section{Numerical Results} \label{sect:num_res}

In this section, we provide two numerical results for the optimal solution obtained in Section~\ref{sect:gen_soln} for $n= 8$. For these results, we consider the update arrival rates at the source $\lambda_i  = bq^i$ with $q =0.7$ such that $\sum_{i=1}^{n}\lambda_i = 10$. We take the file request rates at the user $u_i  = d r^i$ with $r = 0.8$ such that $\sum_{i=1}^{n}u_i = 20$. Finally, we take the file transmission rates at the source $s_i  = h p^i$ with $p = 1.25$ such that $\sum_{i=1}^{n}s_i = 3$.   

\begin{figure}[t]
	\centerline{\includegraphics[width=0.69\columnwidth]{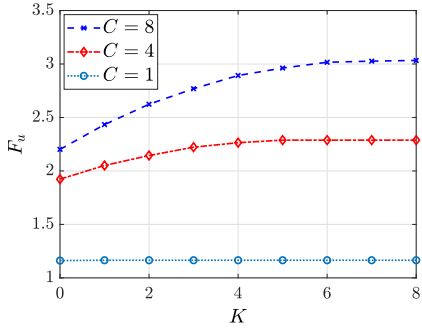}}
	\vspace{-0.35cm}
	\caption{Total freshness of the user $F_u$ with respect to the cache capacity $K$ when the total cache update rate is $C = 1,4,8$.}
	\label{Fig:sim1}
	\vspace{-0.55cm}
\end{figure}

In the first example, we increase the cache capacity $K$ from $0$ to $n$ when the total update rate at the cache is $C = 1,4,8$. We observe in Fig.~\ref{Fig:sim1} that when the total cache update rate is small, i.e., when $C=1$, increasing the cache capacity $K$ does not improve the freshness of the user much, i.e., $F_u$ stays constant for $K\geq 1$. As the total cache update rate $C$ is too low, if a file is stored at the cache, the user gets obsolete versions of the file most of the time. Thus, we observe that even though the cache capacity is high, the optimal policy is to cache only one file. In this case, the cache mostly forwards the update requests from the user to the source, i.e., the cache behaves like a relay node. When the total cache update rate increases, i.e., when $C=4$, we observe in Fig.~\ref{Fig:sim1} that increasing the cache capacity $K$ increases the user freshness up to $K=5$ and does not improve it for $K>5$. Similarly, when $C=8$, we observe in Fig.~\ref{Fig:sim1} that the user freshness increases with the cache capacity. In this case, as the total cache update rate is high enough, the optimal policy is to cache every file.

In the second example, we consider the same system as in the first example, but we take $K=n$ and find the optimal caching status for each file $k_i$ and the corresponding file update rates at the cache $c_i$. When $C=1$, the optimal policy is to cache only the $6$th file, i.e., $k_6=1$ and $k_i =0 $ for $i\neq6$. When $C=4$, the optimal caching status is $k_i = 1$, for $i=3,4,5,6,7$ and $k_i =0$, otherwise. Thus, we observe that the files that change too fast at the source are not cached. Furthermore, as the file transmission rate of the $8$th file is too high, we see that the $8$th file is not cached. When $C=8$, it is optimal to cache every file, i.e.,  $k_i = 1$ for all $i$. Thus, when the total cache update rate is high enough, the optimal policy is to cache every file as caching helps user to avoid the transmission time between the source and the user. However, when the total cache update rate is limited, the optimal policy is not to cache the files that are frequently updated at the source or the files that have smaller transmission times.

\begin{figure}[t]
	\begin{center}
		\subfigure[]{%
			\includegraphics[width=0.493\linewidth]{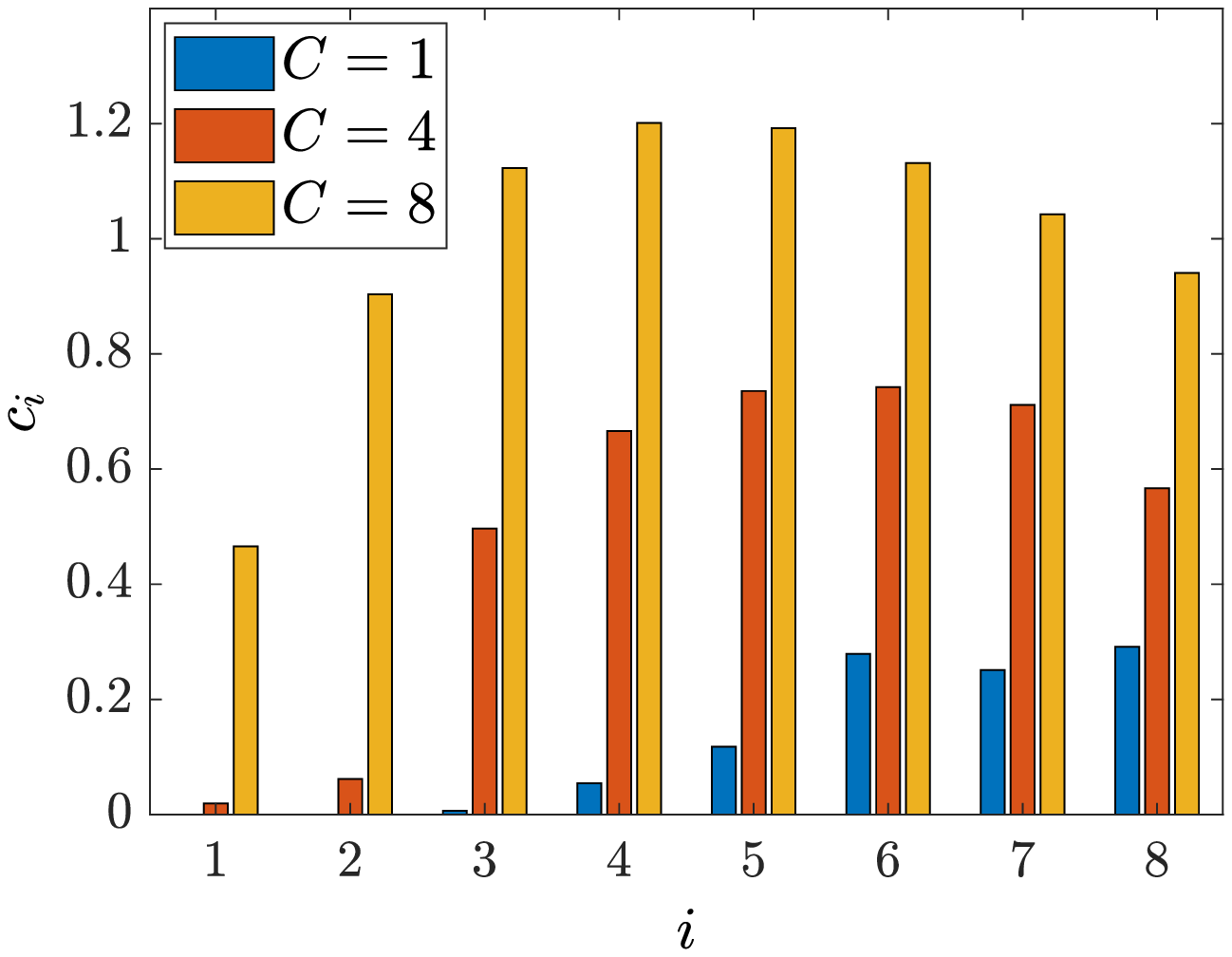}}
		\subfigure[]{%
			\includegraphics[width=0.493\linewidth]{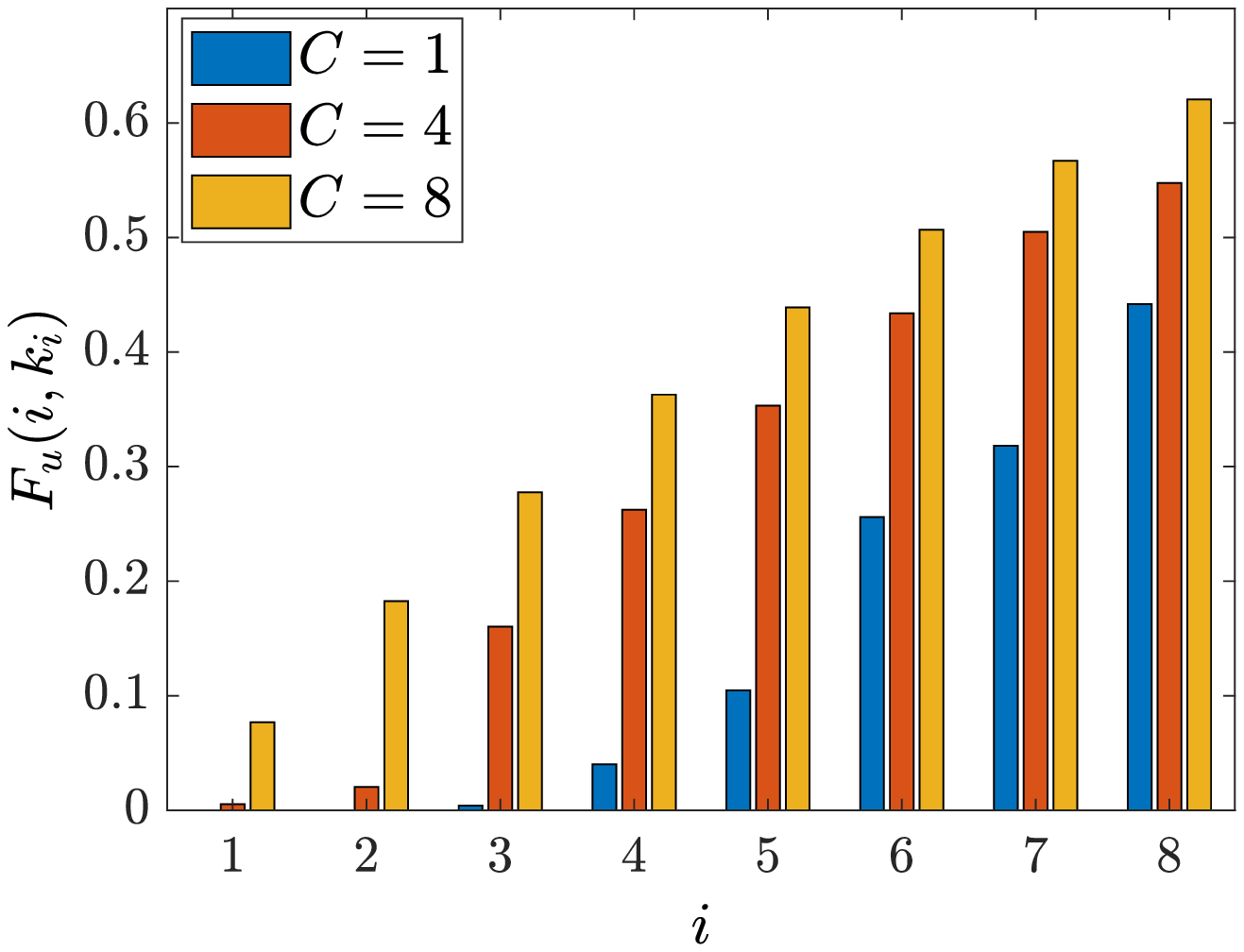}}
	\end{center}
	\vspace{-0.4cm}
	\caption{(a) Update rate allocation at the cache for each file, and (b) the corresponding freshness $F_{u}(i,k_i)$, when $C= 1,4,8$.}
	\label{Fig:sim2}
	\vspace{-0.5cm}
\end{figure}

The optimal file update rate of the cache $c_i$ is shown in Fig.~\ref{Fig:sim2}(a). When $C=1$, i.e., when the total cache update rate is too small, the first two files which are updated at the source most frequently are not updated by the cache, i.e., $c_1=c_2 =0$. Furthermore, we observe in Fig.~\ref{Fig:sim2}(a) that the file update rates at the cache initially increase with the file indices up to $i =6$ when $C=4$ and up to $i=4$ when $C=8$, and then decrease for the remaining files. The freshness of the files at the user $F_u(i,k_i)$ is shown in Fig.~\ref{Fig:sim2}(b). We see in Fig.~\ref{Fig:sim2}(b) that the files that change slowly at the source have higher file freshness at the user even though the file update rates at the cache get lower. We observe that increasing the total cache update rate $C$ improves the freshness of the files. However, the freshness improvement on the rapidly changing files is higher than the others.                             
             
\bibliographystyle{unsrt}
\bibliography{IEEEabrv,lib_v1_melih}

\begin{thebibliography}{10}

\bibitem{Cho03}
J.~Cho and H.~Garcia-Molina.
\newblock Effective page refresh policies for web crawlers.
\newblock {\em ACM Transactions on Database Systems}, 28(4):390--426, December
  2003.

\bibitem{Kolobov19a}
A.~Kolobov, Y.~Peres, E.~Lubetzky, and E.~Horvitz.
\newblock Optimal freshness crawl under politeness constraints.
\newblock In {\em ACM SIGIR Conference}, July 2019.

\bibitem{Kaul12a}
S.~K. Kaul, R.~D. Yates, and M.~Gruteser.
\newblock Real-time status: How often should one update?
\newblock In {\em IEEE Infocom}, March 2012.

\bibitem{Costa14}
M.~Costa, M.~Codrenau, and A.~Ephremides.
\newblock Age of information with packet management.
\newblock In {\em IEEE ISIT}, June 2014.

\bibitem{Soysal19}
A.~Soysal and S.~Ulukus.
\newblock Age of information in {G/G/1/1} systems: Age expressions, bounds,
  special cases, and optimization.
\newblock May 2019.
\newblock Available on arXiv: 1905.13743.

\bibitem{Gao12}
W.~Gao, G.~Cao, M.~Srivatsa, and A.~Iyengar.
\newblock Distributed maintenance of cache freshness in opportunistic mobile
  networks.
\newblock In {\em IEEE ICDCS}, June 2012.

\bibitem{Yates17b}
R.~D. Yates, P.~Ciblat, A.~Yener, and M.~Wigger.
\newblock Age-optimal constrained cache updating.
\newblock In {\em IEEE ISIT}, June 2017.

\bibitem{Kam17b}
C.~Kam, S.~Kompella, G.~D. Nguyen, J.~Wieselthier, and A.~Ephremides.
\newblock Information freshness and popularity in mobile caching.
\newblock In {\em IEEE ISIT}, June 2017.

\bibitem{Zhong18c}
J.~Zhong, R.~D. Yates, and E.~Soljanin.
\newblock Two freshness metrics for local cache refresh.
\newblock In {\em IEEE ISIT}, June 2018.

\bibitem{Zhang18}
S.~Zhang, J.~Li, H.~Luo, J.~Gao, L.~Zhao, and X.~S. Shen.
\newblock Towards fresh and low-latency content delivery in vehicular networks:
  An edge caching aspect.
\newblock In {\em IEEE WCSP}, October 2018.

\bibitem{Tang19}
H.~Tang, P.~Ciblat, J.~Wang, M.~Wigger, and R.~D. Yates.
\newblock Age of information aware cache updating with file- and age-dependent
  update durations.
\newblock September 2019.
\newblock Available on arXiv: 1909.05930.

\bibitem{Yang19a}
L.~Yang, Y.~Zhong, F.~Zheng, and S.~Jin.
\newblock Edge caching with real-time guarantees.
\newblock December 2019.
\newblock Available on arXiv:1912.11847.

\bibitem{Sun17b}
Y.~Sun, Y.~Polyanskiy, and E.~Uysal-Biyikoglu.
\newblock Remote estimation of the {Wiener} process over a channel with random
  delay.
\newblock In {\em IEEE ISIT}, June 2017.

\bibitem{Arafa20a}
A.~{Arafa}, J.~{Yang}, S.~{Ulukus}, and H.~V. {Poor}.
\newblock Age-minimal transmission for energy harvesting sensors with finite
  batteries: Online policies.
\newblock {\em IEEE Transactions on Information Theory}, 66(1):534--556,
  January 2020.

\bibitem{Bastopcu18}
M.~Bastopcu and S.~Ulukus.
\newblock Age of information with soft updates.
\newblock In {\em Allerton Conference}, October 2018.

\bibitem{Buyukates18c}
B.~Buyukates, A.~Soysal, and S.~Ulukus.
\newblock Age of information scaling in large networks.
\newblock In {\em IEEE ICC}, May 2019.

\bibitem{Bastopcu20a}
M.~Bastopcu and S.~Ulukus.
\newblock Who should {Google} {Scholar} update more often?
\newblock In {\em IEEE Infocom}, July 2020.

\bibitem{Bastopcu20d}
M.~Bastopcu and S.~Ulukus.
\newblock Information freshness in cache updating systems.
\newblock April 2020.
\newblock Available on arXiv:2004.09475.

\bibitem{Yates14}
R.~D. Yates and D.~J. Goodman.
\newblock {\em Probability and Stochastic Processes}.
\newblock Wiley, 2014.

\bibitem{Boyd04}
S.~P. Boyd and L.~Vandenberghe.
\newblock {\em Convex Optimization}.
\newblock Cambridge University Press, 2004.

\end{thebibliography}
\end{document}